\documentclass[conference]{IEEEtran}
\IEEEoverridecommandlockouts
\usepackage{cite}
\usepackage{amsmath,amssymb,amsfonts}
\usepackage{algorithmic}
\usepackage{graphicx}
\usepackage{textcomp}
\usepackage{xcolor}
\usepackage{balance}
\usepackage[italian,english]{babel}

\def\BibTeX{{\rm B\kern-.05em{\sc i\kern-.025em b}\kern-.08em
    T\kern-.1667em\lower.7ex\hbox{E}\kern-.125emX}}
\begin{document}

\title{Towards a Scaled IoT Pub/Sub Architecture for 5G Networks: the Case of Multiaccess Edge Computing
}

\author{\IEEEauthorblockN{Alessandro E. C. Redondi}
\IEEEauthorblockA{\textit{DEIB} \\
\textit{Politecnico di Milano}\\
Milan, Italy \\
\{name.surname\}@polimi.it}
\and
\IEEEauthorblockN{Andr\'es Arcia-Moret}
\IEEEauthorblockA{\textit{Computer Laboratory} \\
\textit{University of Cambridge}\\
Cambridge, UK \\
andres.arcia@cl.cam.ac.uk
}
\and
\IEEEauthorblockN{Pietro Manzoni}
\IEEEauthorblockA{\textit{DISCA} \\
\textit{Universitat Politecnica de Valencia}\\
Valencia, Spain \\
pmanzoni@disca.upv.es
}
}

\maketitle

\begin{abstract}
	The vision of the Internet of Thing is becoming a reality and novel communications technologies such as the upcoming 5G network architecture are designed to support its full deployment. In this scenario, we discuss the benefits that a publish/subscribe protocol such as MQTT or its recently proposed enhancement MQTT+ could bring into the picture. However, deploying pub/sub brokers with advanced caching and aggregation functionalities in a distributed fashion poses challenges in protocol design and management of communication resources. In this paper, we identify the main research challenges and possible solutions to scale up a pub/sub architecture for upcoming IoT applications in 5G networks, and we present our perspective on systems design, optimisation, and working implementations. 
\end{abstract}

\begin{IEEEkeywords}
IoT, Pub/Sub, 5G, Multiaccess Edge Computing
\end{IEEEkeywords}

\section{Introduction}\label{sec:introduction}
The Internet of Things (IoT) is becoming a reality, and in the last few years we have indeed witnessed to an enormous growth of technologies designed for its wide and capillary implementation. In particular, many efforts have been made in order to design communication solutions adapted to the specific requirements\footnote{in primis, the power consumption} of IoT devices. Such efforts produced a great variety of different communication technologies tailored to low-power devices, ranging from short-range solutions (IEEE 802.15.4, Bluetooth Low Energy) to dedicated long-range cellular-like networks (LoRa/LoRaWAN, Sigfox, Ingenu). Similarly, other efforts have been also made to adapt traditional mobile cellular networks to machine-type communication typical of the IoT, and solutions like LTE-M or NB-IoT are already available from cellular operators~\cite{cesana2017iot}.

Alternatively, it is expected that the advent of the 5th Generation (5G) of mobile cellular networks will boost tremendously the development and implementation of large scale, city-wide IoT applications. In this respect, two main 5G innovation pillars have been designed precisely for accommodating IoT requirements: massive Machine Type Communication (mMTC) and Multi-Access Edge Computing (MEC). The former will enable connection densities in the order of 10$^6$ low-power devices per square kilometre; while the latter will enable (serverless) distributed computing at the edge of the network, opening it to applications and services from third parties. 

In MEC scenario, we argue that different protocol standards such as MQTT, MQTT+ and distributed orchestration of brokers, will facilitate the development of large-scale interconnected IoT systems. Referring in particular to the all-IP 5G network infrastructure, our discussion concentrates on the higher layers of the TCP/IP stack. Regarding the transport layer, although it is arguable \cite{gomez_tcp_2018} solutions based on (lower overhead) UDP seem to better suit the IoT scenario rather than TCP. As for the application layer, two main communication paradigms are available: Representational State Transfer (REST) and publish/subscribe. HTTP (or its lightweight version COAP~\cite{bormann2012coap}) and MQTT (or its enhanced version MQTT+~\cite{giambona2018mqtt+}) are excellent examples of the two approaches: although both have reached a certain popularity, it is still unclear which one will become a preferred and widely adopted solution in the IoT world.

In this paper, we draw a picture of the future 5G-enabled IoT particularly focusing on the application layer, while reconciling  several architectural and protocol related concepts, and identifying operational meeting points among different research areas. In the particular case of Multi-access Edge Computing, we take a position in favour of publish/subscribe approaches at the application layer and propose MQTT-based approaches as candidates for becoming preferred solutions. We identify the main challenges of this vision and propose possible solutions considering past, present and future approaches.

\section{MEC: enabling IoT in 5G networks}
Multi-access Edge Computing (MEC) is identified as one of the key technologies required to support low latency for future services in 5G networks. The main idea is to bring computational power, storage resource and services typical of nowadays cloud infrastructures to the edge of the network, at close proximity to the users. By doing this latency is greatly reduced, as well as the amount of traffic to be managed by the core network. MEC use case examples include computation offloading, distributed content delivery and caching, web performance enhancements and, of course, IoT applications.

Regarding the latter, MEC technologies are envisioned to work as IoT gateways, facilitating the management of data in close proximity to their sources, providing computational and storage resources as well as processing, aggregation and filtering functionalities~\cite{salman2015edge,porambage2018survey}.
MEC platforms will be offered and deployed by the network operator at multiple locations (e.g., at the base stations, at cell aggregations sites or at multi-RAT aggregation points), and will be also made open to authorised third parties such as application developers and content providers~\cite{mach19mobile}. Motivated by traffic off-loading and considerable reductions in latency, very recently, the major cloud service providers have started working on edge solutions to move part of their services closer to the final users: Amazon AWS Greengrass/Lambda, Google IoT Edge, IBM Watson Edge Analytics and Microsoft Azure IoT Edge can be intended as efforts of such companies to prepare products for the upcoming 5G network architecture based on MEC.

\begin{figure}[t!]
	\centering
    \includegraphics[width=0.8\columnwidth]{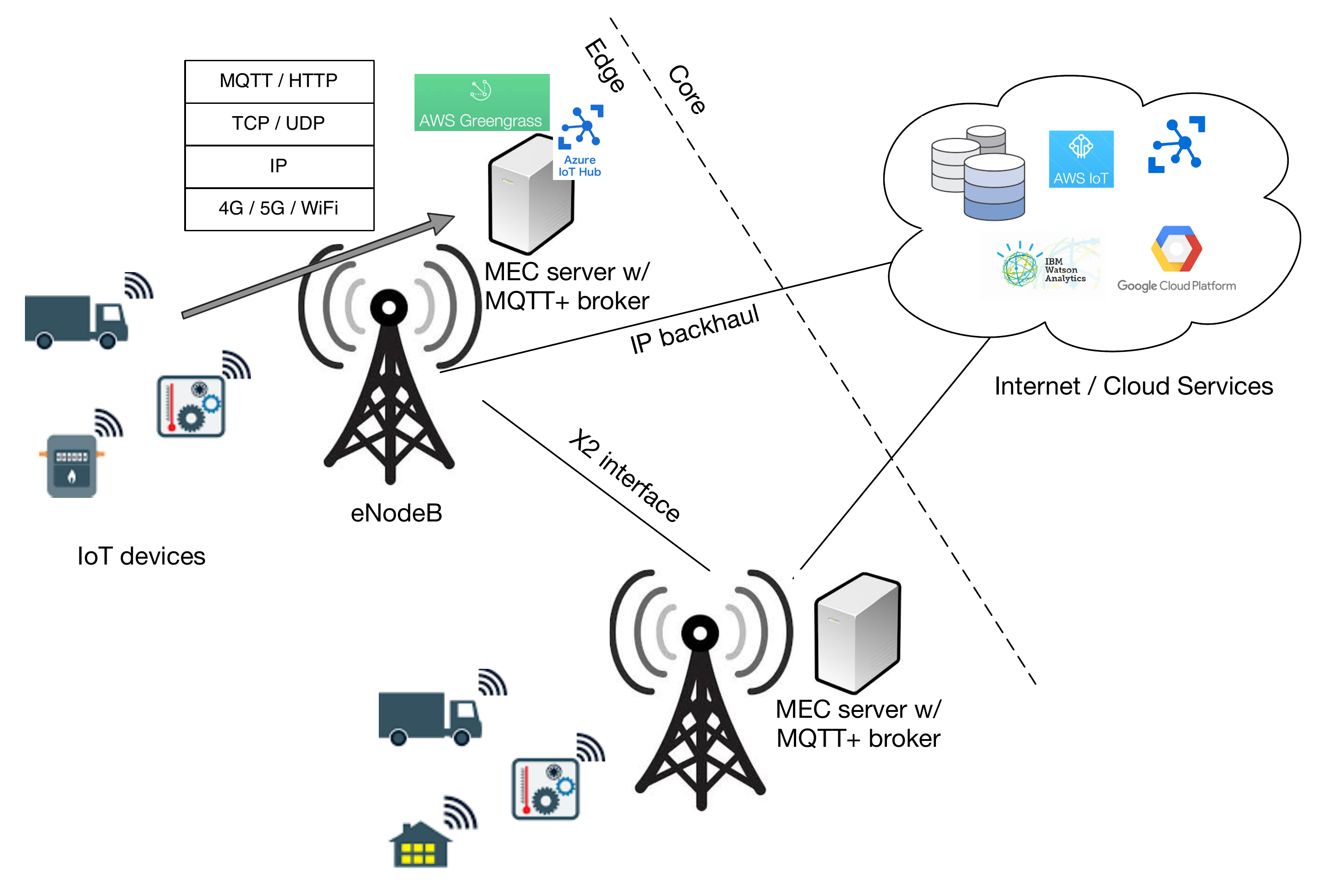}
    \caption{IoT scenario supported by MEC-enabled 5G architecture}
    \label{fig:architecture}
	\vspace{-1em}
\end{figure}

Figure \ref{fig:architecture} briefly illustrates the scenario: IoT devices may be served with different types of connectivity (including WiFi) from the 5G base stations, and communicate through IP and TCP or UDP with the MEC servers and with the Internet, where traditional cloud services are located. As in legacy LTE mobile networks, different base stations may be directly connected to each other through X2 interfaces, facilitating tasks such as devices handovers. At the application layer, we observer a growing dichotomy between RESTful or pub/sub approaches. Indeed, the aforementioned four major edge computing services offer either one or both approaches for connecting IoT devices: Amazon, Google and IBM offer HTTP/HTTPS and MQTT interfaces while Microsoft Azure IoT Edge supports only pub/sub protocols (MQTT or AMQP).

\section{REST or PUB/SUB in IoT? COAP vs MQTT}\label{sec:MQTT+}
\subsection{COAP vs MQTT}
HTTP is the most popular application layer protocol in the Internet ecosystem, and does its job efficiently. Therefore, when designing the application layer of the Internet of Things, researchers tried to adapt the REST approach of HTTP to resource constrained devices. The efforts resulted in the Constrained Application Protocol (COAP), standardised in 2014 by the IETF. COAP, based on UDP, provides the same set of primitives of HTTP (GET, PUT, POST, DELETE) with reduced complexity. COAP-enabled IoT devices hold data and measurements in form of resources, identified with URIs, and act as servers. Clients interested in such measurements access them in a standard request-response fashion. However, such a pull-based approach does not fit well the majority of IoT application scenarios, where devices perform measurements autonomously and transmit them to a central collection point. To overcome this issue and avoid the collection point to continuously poll a resource, COAP provides an observation mode. A client (the central collection point) registers to a resource state on an IoT device and gets notified each time it changes. Although COAP is a reference protocol for low-power devices and its implementation is available for several programming languages, to date it is not taken into consideration by any of the major cloud and edge platform services players\footnote{To the best of our knowledge, the only IoT-related initiative based on COAP is the Open Connectivity Foundation (OCF), which mostly targets the smart home scenario and short range communication technologies rather than cellular ones.}. Therefore, we do not expect it to be used in at least the first rollout of 5G MEC solutions.

Conversely, MQTT is living its greatest period of popularity since its proposal in 1999. Standardised by OASIS in 2014, this lightweight publish/subscribe protocol is practically becoming the standard de-facto in M2M and IoT applications. As a matter of fact, all major cloud platforms (e.g., Amazon AWS, Microsoft Azure, IBM Watson) expose their IoT services through MQTT. The reasons of such popularity derive from MQTT’s incredible simplicity at the client-side, which nicely fits in resource-constrained applications, yet supporting reliability and several degrees of quality of service (QoS). MQTT is based on the publish/subscribe communication pattern, and all communications between nodes are made available via a broker. The broker accepts messages published by devices on specific topics and forwards them to clients subscribed to those topics, ultimately controlling all aspects of communication between devices.
 
\subsection{MQTT+}
Recently, the MQTT protocol has received a lot of attention from the research community. In particular, we mention here MQTT+~\cite{giambona2018mqtt+}, a version that nicely fits with the 5G/MEC scenario under consideration. An MQTT+ broker provides all functionalities of legacy MQTT, but can also perform advanced operations such as spatio/temporal data aggregation, filtering and processing. Such operations are triggered by specific topics, as demonstrated below:
\begin{itemize}
	\item{\textbf{Data filtering:} MQTT+ allows a client to perform a rule-based subscription using ad-hoc prefix operators. As an example, a client subscribing to \verb|$GT;value/topic/| will receive only messages published on \verb|topic| that contain a value greater than \verb|value|. Other comparison operators are defined, such as lower than (\verb|$LT|), equals or not equals (\verb|$EQ| / \verb|$NEQ| ) and contains (\verb|$CONTAINS|)}.
	\item{\textbf{Temporal aggregation:} MQTT+ allows a client to subscribe to certain temporal aggregation functions of a topic, using the format \verb|$<TIME><OP>/topic|, with \verb|OP| = \verb|{COUNT,SUM,AVG,MIN,MAX}| and \verb|TIME| = \verb|{DAILY,HOURLY,QUARTERHOURLY}|. This allows a client to obtain e.g., the daily count of messages on a certain topic. The MQTT+ broker handles all operations internally by caching values and computing aggregates for specific time intervals.}
	\item{\textbf{Spatial aggregation:} MQTT+ provides a client with the possibility to subscribe to multiple topics at once by using a single-level (+) or multi-level (\#) wildcard. However, a client may be interested in aggregating such topics at once: MQTT+ allows for this possibility. A client may subscribe to \verb|$<OP>/topic/|, where \verb|OP| can assume the same values defined for temporal aggregation and \verb|/topic/| contains one or more wildcards. By doing this a client can obtain, e.g., the average or the sum of the values published by different sensors.}
	\item{\textbf{Data processing:} Beside simple temporal and spatial aggregation, MQTT+ allows a client to subscribe to processing operations executed by the broker on multimedia data (audio, images and video) published by sensors. The broker advertises its processing capabilities under a special topic (e.g., \verb|$SYS/capabilities/|). Specific operators (such as the \verb|$CNTPPL| prefix to count people) trigger the broker to run specific algorithms and to return the result to the subscriber. Clients may use such capabilities to obtain processed information from the raw data, avoiding the need to perform processing themselves. As an example, a client may subscribe to the \verb|$CNTPPL/camera_id| to obtain the number of people contained in the images published on the \verb|camera_id| topic. When the MQTT+ broker is implemented on a MEC server run by one of the major cloud operators, such advanced capabilities may be provided by one of the existing cloud processing tools (e.g., Amazon Rekognition, Google Vision).}
	\item{\textbf{Composite subscriptions:} One of the strengths of MQTT+ is the capability of allowing composite subscriptions by properly chaining the operators introduced so far, thus enabling even more advanced functions. Indeed, MQTT+ supports spatio-temporal aggregations, spatio-temporal aggregation of processed data and even rule-based spatio-temporal aggregation. To give a concrete example, a subscription to \verb|$DAILAVG$CNTPPL/camera_id|} triggers the broker to count the number of people contained in all images published on the \verb|camera_id| topic, returning to the subscriber its daily average.
\end{itemize}

In the next section we discuss the research challenges of operating an MQTT+ broker in a MEC server, in the context of a 5G network.

\section{MQTT+ on MEC: research challenges}


\begin{figure}[t!]
	\centering
    \includegraphics[width=0.8\columnwidth]{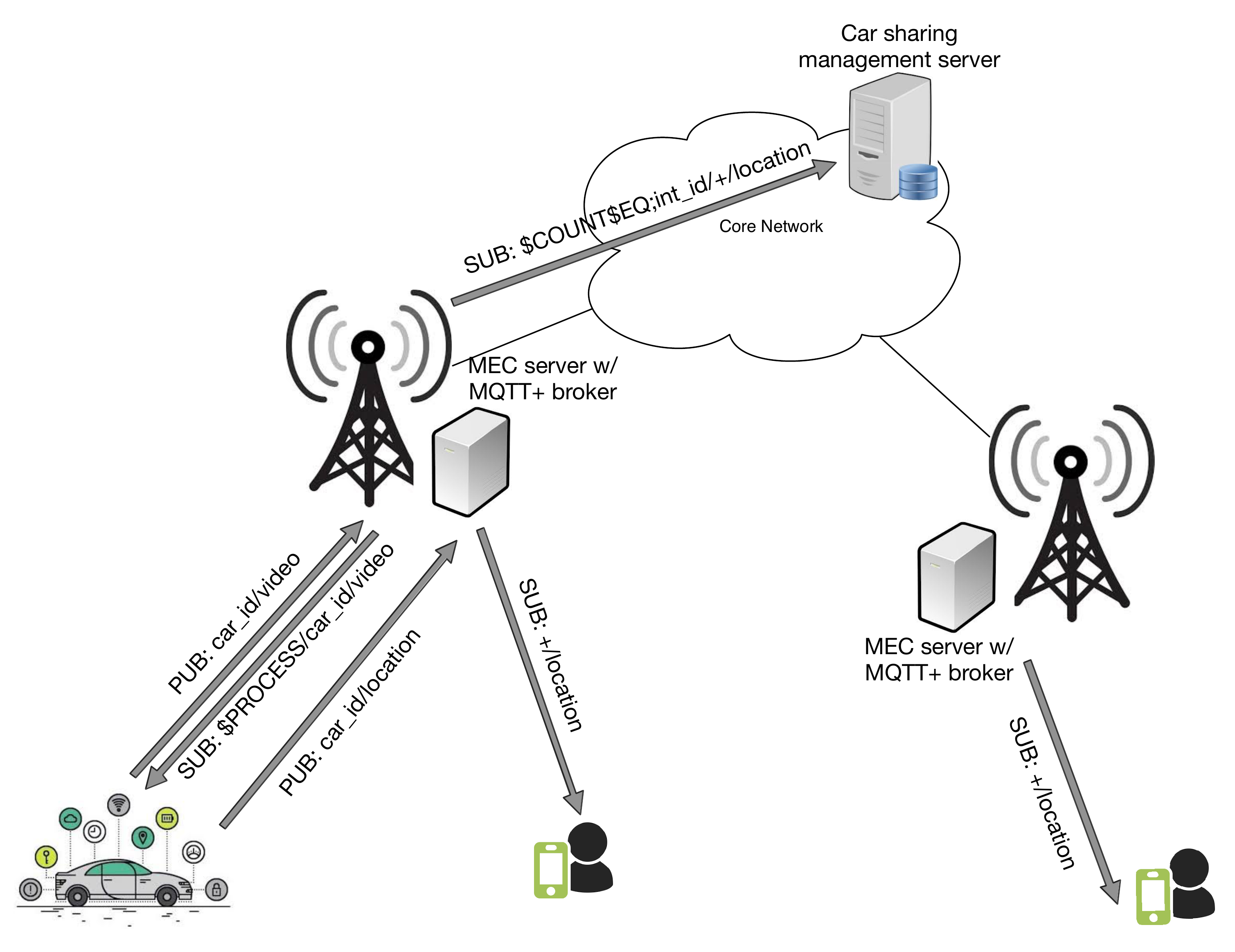}
    \caption{IoT use case: vehicle sharing}
    \label{fig:architecture_2}
	\vspace{-1em}
\end{figure}

We take as reference the example use-case of a vehicle data sharing (cars or bicycles) system implemented in a 5G-enabled smart city, similar to the ones already deployed in many cities worldwide. An architectural sketch of the system is depicted in Figure \ref{fig:architecture_2}: shared vehicles are information producers and periodically publish to the system meaningful information such as their location, service status and data retrieved from a plethora of sensors (both vehicle-related and environmental-related). Such information are received at an MQTT+ broker installed on a MEC server at the closest base station (or multi-RAT aggregation point) and forwarded to two types of data subscribers: local and global subscribers. Local subscribers are data consumers with low-latency requirements, located in close proximity of the producers (e.g., a person willing to search for the closest vehicle), or which even coincide with the producers themselves. The latter case is represented by all those applications of MEC-enabled offloading, such as intense image/video processing to be performed on multimedia streams coming from in-car cameras (e.g., augmented reality). Such type of processing could be enabled by the advanced broker functionalities provided by MQTT+: in the example shown in Figure \ref{fig:architecture_2} a connected car publishes a video from one of its camera sensors on the topic \texttt{car\_id/video} and subscribes to an advanced service (e.g. augmented reality) on the local MEC broker using the MQTT+ syntax \texttt{\$PROCESS/car\_id/video}. Conversely, global subscribers are consumers located far away from the producers, such as generic users, control operational points, traffic information services and so on. Note that such global subscribers may be more interested in aggregated information rather than raw data (e.g., the amount of cars flowing through an intersection every 5 minutes), again motivating the adoption of an advanced protocol such as MQTT+. As an example, the car sharing a management server in Figure \ref{fig:architecture_2} subscribes to the advanced topic \texttt{\$COUNT\$EQ;int\_id/+/location} to directly obtain the count of all cars passing through a particular intersection \texttt{int\_id}.
This scenario shares many similarities with the work presented in \cite{manzoni_proposal_2017}, where \textit{content islands} of things operated by a pub/sub architecture were organised using local and global topics to differentiate how to manage their publications. In this work we observe that the presence of local and global subscribers, possibly in a mobile scenario, dictates several different requirements on the pub/sub architecture, resulting in several research challenges which are listed in the following.

\subsection{Automatic broker discovery}
MQTT, and by inheritance MQTT+, require client devices to know the IP address of the broker (or the load balancer in case of clustered brokers) in order to connect to it. In a mobile scenario, where clients devices move from the area covered by one MEC server to another, it is important to establish automatic and dynamic procedures for disseminating the broker IP address to the clients. Solutions such as Zeroconf\cite{steinberg2005zero} may be adapted in order to facilitate the task.

\subsection{Broker vertical clustering}
A general issue of MQTT is the central role of the broker. Indeed, if not dimensioned properly, the broker can become the bottleneck and result in a single point of failure, causing the local client devices to be unable to communicate with the network. This problem is even more important with MQTT+ brokers, which require additional resources for data aggregation, filtering and processing. Recent MQTT implementations have the possibility to be vertically clustered (i.e., implemented on several virtual machines on the same physical hardware) to provide some sort of reliability in case of broker failures or overloads. Often, a load balancer is used as a single point of entry for all communications: this creates a single logical broker from the perspective of clients and provides some sort of reliability and vertical scalability~\cite{jutadhamakorn2017scalable,sen2018highly}. However, such setups are either static and cumbersome to deploy or dynamically implement autoscaling in a central cloud service~\cite{gascon2015dynamoth}.  Given the limited amount of resources that will be available at the MEC, solutions where additional brokers may be dynamically created or shut down according to the local load conditions become of primary importance, as well as the development of accurate prediction models for resource provisioning~\cite{tabatabai2017managing}. For MQTT+, this means developing optimisation and design techniques for accepting/rejecting a subscription to an intensive data processing task or to move the corresponding computation elsewhere (e.g. from the MEC to the cloud).

\subsection{Broker distribution and horizontal clustering}
In the scenario depicted by the upcoming 5G architecture, multiple MEC servers, rather than a single central cloud-based server, are deployed in close proximity to the final users. Interconnecting such nodes together is crucial for realising the vision of edge computing, with clear benefits in terms of end latency and use of network resources compared to a cloud-based approach. The interconnections between brokers installed on the MEC servers can be realised either with virtual links based on the S1 interface through the core network or by exploiting the X2 interfaces connecting directly different base stations. In both cases, the main challenge is how to distribute efficiently subscriptions and publications from one broker to other brokers, ultimately interconnecting local publishers with global subscribers. The problem is known as distributed event routing, and has received a lot of attention in the past for what concerns generic publish/subscribe architectures~\cite{baldoni2005distributed,martins2010routing}. However, to the best of our knowledge, no off-the-shelves solutions are ready to be used for interconnecting MQTT and MQTT+ brokers in a distributed fashion: in the next Section we propose three different alternatives for solving such a problem.

\section{Approaches for distributing brokers}

\subsection{Static Broker Bridging}
A naive solution to the problem may be the use of bridging, a functionality already present on some MQTT brokers implementations (e.g., Mosquitto\footnote{https://mosquitto.org} and HiveMQ\footnote{http://www.hivemq.com}) which allows a broker B to connect to broker A as a standard client and subscribe to some or all messages published on A. Vice versa, A is subscribed and receives messages published on B. Despite its simplicity, such an option has several drawbacks. First, to avoid message looping (A publishes a message on B, which in turn forwards it on A), such method requires specific prefix to be added to topic description on each broker. Second, such a mechanism is static in nature and do not address mobility or changing resource availability, although some recent work proposed dynamic bridging tables~\cite{rausch2018emma}. Third, topic bridging is basically equivalent to event flooding in distributed pub/sub system, a solution which is known to not scale well in large scale distributed scenarios~\cite{baldoni2005distributed}.

\subsection{Selective Event Routing}
More efficient solutions may be the ones stemming from the works on event routing in distributed pub/sub systems. In particular, rendezvous-based event routing has the potential to solve the scalability issues arising from massive IoT applications. In rendezvous-based system, publishers and subscribers meet each others at specific nodes in the network, known as rendezvous nodes (RN), which are organised in an overlay network topology. Each RN is responsible for (i) storing the subscriptions to a specific topic or subset of topics and (ii) routing any incoming publication to the RN node in charge of such topics (either directly or through some aggregation function). Subscription and publications therefore meet at the RN node which are both mapped to. Mapping between topics and RN nodes is generally performed through the use of hashing functions, which can also be used to balance the load of subscriptions storage and maintenance~\cite{martins2010routing}. While promising, such an approach has two drawbacks: (i) the subscription language is limited by the chosen mapping between subscriptions and RNs and (ii) mobility of publishers is not well managed by a fixed allocation of subscriptions to RNs~\cite{baldoni2005distributed}. Note that, in principle, any MQTT+ broker can host rendezvous functionalities. An interesting research problem is therefore to select which MQTT+ brokers are more suitable to become RNs, based on specific objective functions such as minimising latency.

\subsection{ICN-based approach}
\begin{figure}[t!]
	\centering
    \includegraphics[width=\columnwidth]{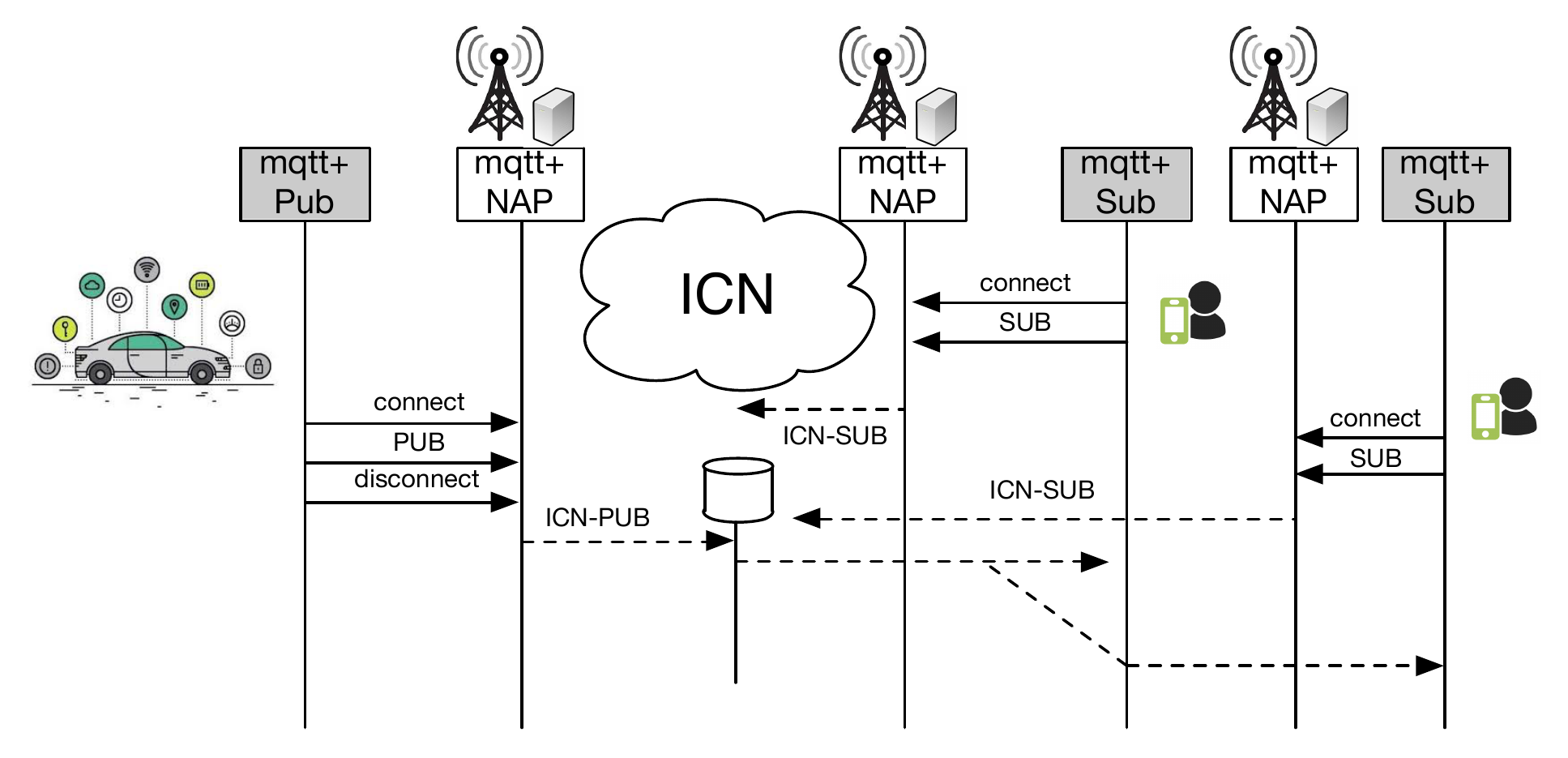}
    \caption{ICN-based routing: the NAP implemented on MQTT+ brokers translate between IP address and ICN names}
    \label{fig:icn-architecture}
	\vspace{-1em}
\end{figure}

As explained before, MQTT+ provides caching of aggregated IoT measurements coming from end devices. Such functionality is useful to deliver essential information to the upper layer services while reducing the amount of data to be managed, ultimately trading off the fine-grained data locations (i.e., the IP address or topic name of each publishing device) with data content (e.g., an aggregation function over the data published). This observation naturally brings into the game the concept of Information Centric Networking (ICN), a novel clean-slate networking paradigm which considers information as the new waist of the Internet communication model. In ICN the focus of the communication becomes \textit{what} it is communicated instead of \textit{where} it is located, i.e., ICN focuses on the naming rather than on the addressing\footnote{See Soch \cite{shoch_note_1978} for further elaboration on this difference}. 
Among the different realisation of the ICN concept, the one which fits best in the reference scenario of this work is POINT~\cite{trossen_ip_2015}, which offers a convenient ICN implementation framework based on a publish/subscribe architecture. The architecture of POINT relies on three complementary network functions: (1) the Topology Manager (TM) for calculating the delivery tree in a one-to-many communication pattern, (2) the Rendezvous Function (RF) to provide the directory and binding service matching up publishers and subscribers (similar to Rendezvous-based Routing) and (3) the Forwarding function that allows the efficient dissemination of information through the use of Forwarding Nodes (FN). FNs are bespoke devices that live alongside the routers in an overlay topology and provide different specific network services such as in-network data aggregation, redundancy elimination and smart caching. As mentioned in Section \ref{sec:MQTT+}, MQTT+ brokers have the right characteristics to act as an interface to a POINT name-based ICN world. The translation function between the IP-based world and ICN is implemented in a Network Attachment Point (NAP), which is also in charge of synchronising all MQTT+ brokers distributed in different parts of the network.
As shown in Fig. \ref{fig:icn-architecture}, we expect island of things to be interconnected, and consequently information and services living in different parts of the network. ICN functions will then be in charge of realising the efficient dissemination across the network connecting the publications and subscriptions to and from the edge of the network. Notice that things are oblivious of the location of the service, since the NAP on the MQTT+ broker takes care of the pertinence of the scope of the message.

\section{Conclusion}

We discussed the challenges associated to the use of a pub/sub protocol such as MQTT+ as enabler of IoT applications in future MEC-enabled 5G networks, and we identified possible future research directions to focus on. Broadening the scope of IoT information dissemination via broker distribution is a key challenge that deserves close attention. It requires the exploration of novel directions on implementation of data managers at the edge. The orchestration strategies for the distribution of these managers over the Internet will also pose further challenges in resource allocation and load balancing. Finally, exploring alternative approaches such as Informaiton Centric Networking for efficient dissemination of information among brokers, promises better tailored communication following many-to-many communication patterns. 



\balance

\section*{Acknowledgment}
Andr\'es Arcia-Moret is under the support of Grant RG90413 NRAG/527.

\bibliographystyle{IEEEtran}
\bibliography{bibfile}

\end{document}